\documentclass[%
 aip,
 amsmath,amssymb,
 reprint,%
]{revtex4-1}

\usepackage{graphicx}
\usepackage{dcolumn}
\usepackage{bm}

\usepackage[utf8]{inputenc}
\usepackage[T1]{fontenc}
\usepackage{mathptmx}

\begin{document}

\preprint{AIP/123-QED}

\title{Inverse Spin Hall effect in ferromagnetic nanomagnet. Dependencies on magnetic field, current and current polarity }

\author{Vadym Zayets}
 \affiliation{National Institute of Advanced Industrial Science and Technology (AIST), Umezono 1-1-1, 
Tsukuba, Ibaraki, Japan}
\author{Andrey S. Mishchenko}
\affiliation{RIKEN Center for Emergent Matter Science (CEMS),  2-1 Hirosawa,  
Wako, Saitama, 351-0198, Japan}

\date{\today}

\begin{abstract}
The measured  Hall angle in a ferromagnetic nanomagnet shows a substantial non-linear dependence on an external magnetic field, which cannot be explained  by adopted mechanisms of the Ordinary and Anomalous (AHE) Hall effects implying
a linear plus constant dependence on the external magnetic field. We suggest that there is an additional non- linear contribution  from the Inverse Spin Hall effect (ISHE). 
The significant contribution of ISHE in a ferromagnet is supported by perfect agreement of experiment 
with a phenomenological theory of ISHE. We observed different dependencies of AHE and ISHE on current suggesting their different thermal dependencies.   
We also observe dependence of the Hall angle of the current polarity which is due to the Spin Hall effect.
\end{abstract}

\maketitle

\section{\label{sec:level1}Introduction}

The Hall effect (HE) is a generation of an electric current perpendicularly to the bias current, 
which flows along  an applied electric field.
The measure of the HE is the Hall angle 
$\alpha_{\mbox{\scriptsize HE}}=\sigma_{xy}/\sigma_{xx}$, which is defined as the  ratio of 
non-diagonal $\sigma_{xy}$ and diagonal $\sigma_{xx}$ conductivities. There are several contributions to the Hall effect in a ferromagnetic metal.
The first considered mechanism of HE is the Ordinary Hall effect (OHE), which is created by the Lorentz force and lineally proportional to the external magnetic field  $\sim \alpha_{\mbox{\scriptsize OHE}} H$, where 
$\alpha_{\mbox{\scriptsize OHE}}$ is the OHE coefficient\cite{Hall1879}.
Another contribution is the Anomalous Hall effect (AHE). 
The AHE  occurs due to the scattering of carriers on the aligned local magnetic moments in a ferromagnet 
and  its contribution is proportional to the total spin of localized d-electrons \cite{PughRostoker1953}. 
The AHE contribution is independent on external magnetic field $H$ provided the local magnetic moments are 
field- independent. This is the case\cite{Gusack} when temperature is not in close vicinity of Curie temperature $T_c$ ($T/{T_c} < 0.99$) and at a moderate magnetic field (H<1T).
The joint contribution of the OHE and AHE to the Hall angle is the sum of the field- independent 
AHE $\sim \alpha_{\mbox{\scriptsize AHE}}$  and the linear OHE  $\sim \alpha_{\mbox{\scriptsize OHE}} H$, which is the prototypical case encountered in plenty 
of magnetic compounds \cite{Sinova}.

The observed nonlinear dependence of HE on the external magnetic field suggests that either the AHE and OHE dependencies are non-liner or there is an additional contribution to the HE. 
As has been shown in Ref.~[\onlinecite{weare}], the non-linear AHE and OHE contributions can be excluded in our specimens  
and the only candidate left is the  Inverse Spin Hall effect (ISHE), which describes the fact that an 
electrical current is created perpendicularly to a flow of  spin-polarized conduction electrons 
\cite{Saitoh,Tinkham, Maekawa}.

The existence of the ISHE in a non-magnetic material has been verified experimentally \cite{r8,Chazalviel1972,Chazalviel1975}. 
In equilibrium the electron gas is not spin-polarized in a non-magnetic material and there is no 
ISHE contribution. 
However, when the spin polarization is externally created, the ISHE contribution can be detected and identified. In experiments of Refs.~[\onlinecite{Chazalviel1972,Chazalviel1975}], the conduction electrons were spin-polarized due to alignment of their spins along an external magnetic field. Their HE contribution was measured by a resonance technique.  In experiment of Ref.~[\onlinecite{r8}] the spin polarization in a paramagnetic AlGaAs/GaAs heterojunction 
was created by circularly-polarized light.
The dependence of the measured Hall angle on the degree of circular polarization and therefore on the spin polarization was clearly detected \cite{r8} confirming the existence of the ISHE contribution. 
  
The possibility of the ISHE contribution in a ferromagnetic metal was addressed experimentally 
only recently \cite{weare}.  
A measurement of the ISHE contribution in a ferromagnetic metal is more difficult,
because the electron gas is spin-polarized even in an equilibrium and, therefore,
the ISHE contribution exists even in equilibrium.
As a result, ISHE can not be pinned down by inducing of spin polarization of otherwise unpolarized 
electron gas.  
A possible way to detect the ISHE in ferromagnetic metal is an modulation of 
the number of spin- polarized electron by an external 
magnetic field \cite{LL,Chazalviel1972,Chazalviel1975,Zay}. 
The nature of the ISHE must inevitably lead to a change of the Hall angle in the external magnetic field. Similar method of identification of the ISHE contribution from a modulation of the spin polarization by an external magnetic field has been used in experiments of  Refs.~[\onlinecite{Chazalviel1972,Chazalviel1975}].
The severe caveat of such approach lies in the fact that the average local magnetic moment can be 
field-dependent as well, which also leads to a field dependence of HE. 
Therefore, it seems that one cannot pin down the ISHE if there is a suspicion that any substantial realignment of 
localized moments can occur like, e.g., in a paramagnet \cite{Maryenko2017}.

Eventually, the idea of detecting the ISHE by a dependence of HE on the external magnetic field can be 
accessible in a ferromagnet if, and only if  the local magnetic moments are appreciably 
independent on the external magnetic field.    
Then, a nanomagnet made of a ferromagnetic metal with the perpendicular magnetic anisotropy 
(PMA) \cite{PMA-review} is just hunted unique object required for such a measurement.  
In such nanomagnet, the magnetic moments are firmly aligned perpendicularly  to the film surface due 
to the strong PMA effect. 
The nano-size of the nanomagnet ensures a one-domain state, in which all localized moments are 
aligned in one direction.
As a result, the AHE contribution becomes essentially independent on external magnetic field
and the joint contribution of the OHE and AHE is strictly a sum of field independent 
AHE $\alpha_{\mbox{\scriptsize AHE}}$  and linear OHE  $\sim \alpha_{\mbox{\scriptsize OHE}} H$
terms.   

In this study we perform field dependent measurements of the HE angle in a FeB nanomagnet with a strong PMA effect.
We develop a phenomenological theory of the ISHE dependence on the external magnetic 
field $H$ which is in perfect agreement with the  experimental dependence of the Hall angle on external magnetic field. 
We also measure the dependence of the HE on the magnitude and direction of the bias current and conclude 
that the observed current direction asymmetry provides an additional argument in favor of the importance 
of the ISHE in ferromagnetic metals.

\section{Experiment and Phenomenological theory of ISHE in a ferromagnetic nanomagnet}

The Hall angle is measured in 1.1-nm-thick FeB grown on $SiO_2/Ta(2.5 nm)$  and covered by $MgO$ using a Hall-bar setup. The width and length of nanomagnet are 800 and 1000 nm, correspondingly. Details of fabrication and measurement are described Ref.~[\onlinecite{weare}].

\begin{figure}[b]
\includegraphics[width=8.5cm]{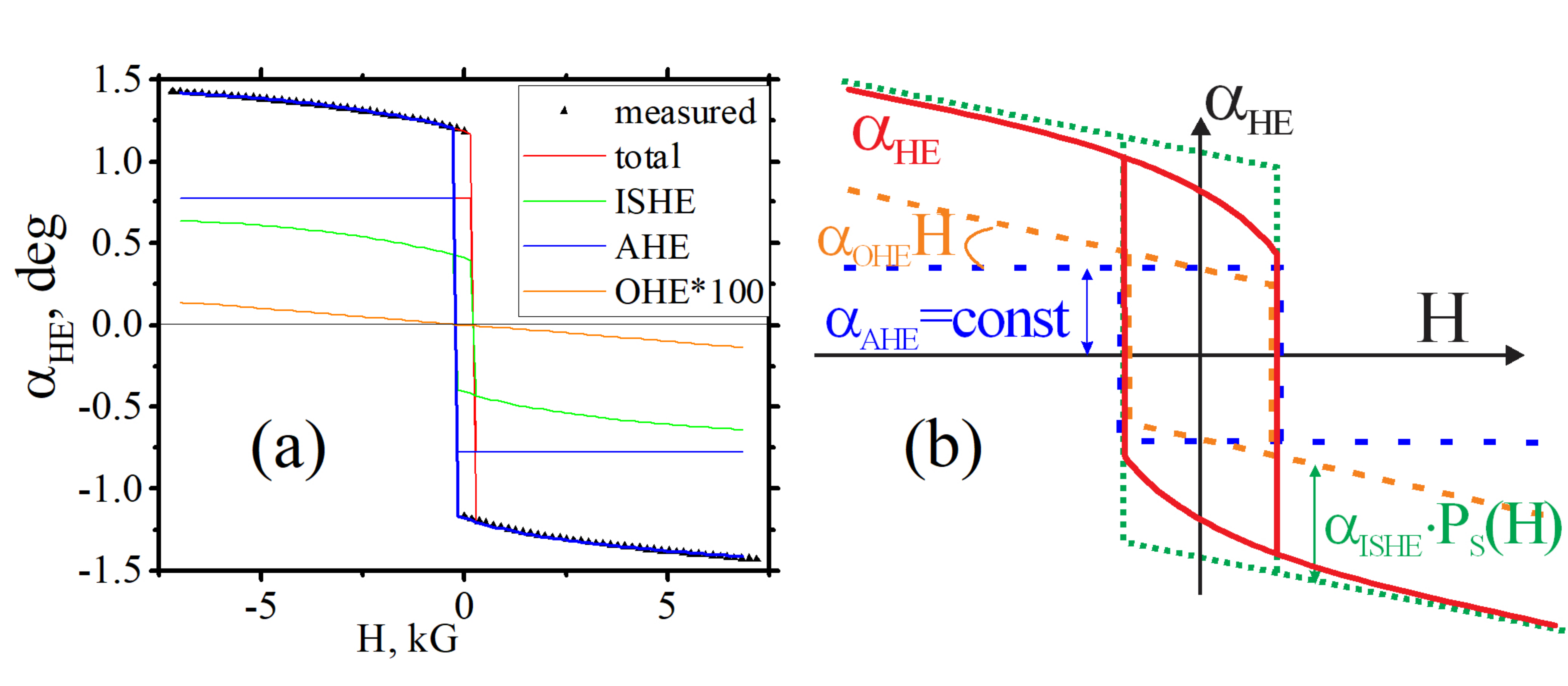}
\caption{\label{fig:fig1}
Hall angle vs perpendicular magnetic field H. 
(a) Experimental data as a sum of three contributions:
triangles is the total measured angle $\alpha_{\mbox{\scriptsize HE}}(H)$, red line is the
sum of contributions, blue/green/orange lines are individual
AHE/ISHE/OHE contributions, respectively. (b) Schematic
plot showing the shape of hysteresis loop for different contributions. 
Solid red line is the measured (total) angle. Dashedblue horizontal line is hypothetical loop in absence of OHE
and ISHE. Slanted dashed orange line is loop in absence of
only ISHE. Dotted green line is the loop in the hypothetical
case when all conduction electrons are spin polarized $P_S=1$ .
}
\end{figure}

We present the Hall angle  $\alpha_{\mbox{\scriptsize HE}}$ as a sum of three terms \cite{weare}. 
\begin{equation}
\alpha_{\mbox{\scriptsize HE}} (H)= 
\alpha_{\mbox{\scriptsize OHE}} H +
\alpha_{\mbox{\scriptsize AHE}} +
\alpha_{\mbox{\scriptsize ISHE}} P_S (H) \; ,
\label{ha}
\end{equation}
where coefficients $\alpha_{\mbox{\scriptsize AHE, OHE,ISHE }}$ are field- independent and the 
function $P_S (H)$ is defined as the spin polarization of the conduction 
electrons in the external filed $H$. 
We show in Figure~\ref{fig:fig1} how $\alpha_{\mbox{\scriptsize HE}} (H)$ can be divided into three contributions. 

The ISHE originates from the spin-dependent scatterings of the conduction electrons when the amount 
of spin-polarised conduction electrons scattered into the left/ right directions are different due to the Spin-Orbit Interaction\cite{weare}.
As a result, the ISHE contribution is linearly proportional to the number of spin-polarized electrons and therefore 
 to the spin polarization   
\begin{equation}
P_S(H) = n_{\mbox{\scriptsize SP}} / \left(  n_{\mbox{\scriptsize SP}}+n_{\mbox{\scriptsize SU}} \right) \; 
\label{p_s}
\end{equation}
in the external magnetic field $H$. This fact can be understood as follows.
All conduction electrons $n_{\mbox{\scriptsize SP}}+n_{\mbox{\scriptsize SU}}$ in Eq.~(\ref{p_s}), 
which participate in charge transport, are divided into the group of the spin-unpolarized electrons
$n_{\mbox{\scriptsize SU}}$ and the group of the spin-polarized electrons $n_{\mbox{\scriptsize SP}}$.
All electrons $n_{\mbox{\scriptsize SP}}+n_{\mbox{\scriptsize SU}}$
contribute to the diagonal component $\sigma_{xx}$ of the conductivity whereas only the spin-polarized 
ones $n_{\mbox{\scriptsize SP}}$ participate in $\sigma_{xy}$. Since the Hall angle is a ratio of the nondiagonal $\sigma_{xy}$ and diagonal $\sigma_{xx}$ conductivities, ISHE contribution to the Hall angle
is proprtional to the spin polarization ${P_S}\left( H \right)$       

The conversion rate between spin-polarised 
$n_{\mbox{\scriptsize SP}}$ and spin-unpolarised $n_{\mbox{\scriptsize SU}}$  
electrons can be calculated as \cite{Za2,weare} 
\begin{equation}
\frac{\partial n_{\mbox{\scriptsize SP}}}{\partial t}  =
\left[
\frac{n_{\mbox{\scriptsize SU}}}{\tau_{M}} - 
\frac{n_{\mbox{\scriptsize SP}}}{\tau_{\mbox{\scriptsize rel}}}
\right] 
+
\frac{n_{\mbox{\scriptsize SU}}}{\tau_{H}}
\; ,
\label{d_n}
\end{equation}
where $1/\tau_{M}$ and $1/\tau_{H}$ are the rates of spin-pumping processes 
$n_{\mbox{\scriptsize SU}} \to n_{\mbox{\scriptsize SP}}$ whereas the rate $1/\tau_{\mbox{\scriptsize rel}}$
describes the spin-relaxation  $n_{\mbox{\scriptsize SP}} \to n_{\mbox{\scriptsize SU}}$
into spin unpolarized states.  
The spin pumping rate $1/\tau_{M}$ is set by spin-dependent scattering on the alligned 
magnetic moments $M$ and $1/\tau_{H}$ describes the rate of the spin-polarization processes caused
by external magnetic field.  
The expression for spin polarization $P_S (H)$ follows from
Eqs.~({\ref{p_s}-\ref{d_n}}) and balance condition $\partial n_{\mbox{\scriptsize SP}}/\partial t=0$,
\begin{equation}
P_S (H) = 
\frac
{P_s^{(0)} + (H/H_S)    }
{1+(H/H_S)  }
\; ,
\label{p_s_h2}
\end{equation}
where the spin-polarization in the absence of the external magnetic filed is
$P_s^{(0)} =  \tau_{\mbox{\scriptsize rel}} / 
\left(  \tau_{\mbox{\scriptsize rel}} + \tau_{M}\right)$
and $H_s$ is the scaling relaxation magnetic field ${H_S} = 1/(\varepsilon  \cdot {\tau _{rel}})$, which is determined by the relation between the spin-depolarization relaxation rate 
$1 / \tau_{\mbox{\scriptsize rel}}$ and magnetic field alignment rate per field unit $\varepsilon$ \cite{Zay}.

It was shown \cite{weare} that the experimental variation of $\alpha(H)$ vs. $H$ can be perfectly described 
in terms of the relation (\ref{ha}) depending on five fit parameters,
namely $\alpha_{\mbox{\scriptsize OHE}}$, $\alpha_{\mbox{\scriptsize AHE}}$, 
$\alpha_{\mbox{\scriptsize ISHE}}$, $H_S$, and $P_s^{(0)}$.
Therefore, we can conclude on the essential role of the ISHE in HE in feromagnets and consider 
the summary in Fig.~\ref{fig:fig1} as an illustration of the structure of the HE in ferromagnets.

\section{Current dependence of the Hall effect}

\begin{figure}
\includegraphics[width=8.7cm]{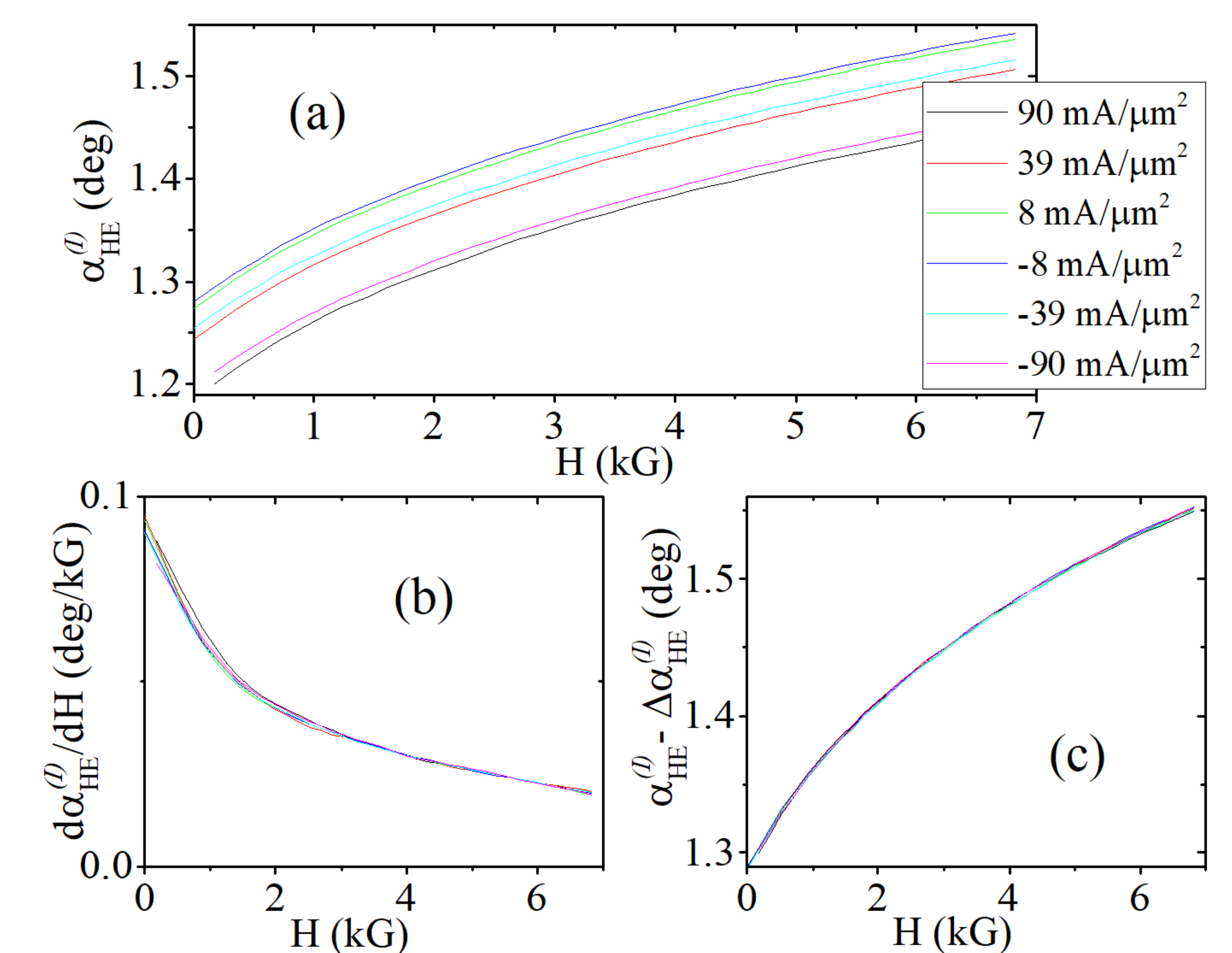}
\caption{\label{fig:fig2}
External magnetic field $H$ dependence of
(a) Hall angle $\alpha_{HE}^{(I)}$, (b) its first derivative over $H$, and 
sums of $\alpha_{HE}^{(I)}$ and fitting $H$-independent  constant 
$\Delta \alpha_{\mbox{\scriptsize HE}}^{(I)}$ for different current densities $I$.
}
\end{figure}

Figure~\ref{fig:fig2}a,b shows the Hall angle $\alpha_{\mbox{\scriptsize HE}}^{(I)}$ and its first derivative measured at a different 
current density $I$.  
The dependence of the Hall angle $\alpha_{\mbox{\scriptsize HE}}^{(I)}$ on current is clear and substantial. In contrast, the current dependency of its first derivative is weak. It indicates a substantial current dependency of AHE, but a weak dependency of ISHE on current. This fact can be understood as follows. The AHE is independent of H and  it contributes only to $\alpha_{\mbox{\scriptsize HE}}^{(I)}$, but not to $d\alpha_{\mbox{\scriptsize HE}}^{(I)}/dH$. In contrast, the non-linear ISHE  contributes to both. The change of $\alpha_{\mbox{\scriptsize HE}}^{(I)}$ without corresponded change of $d\alpha_{\mbox{\scriptsize HE}}^{(I)}/dH$ can occurs only due to a change of AHE. 

 As was shown in Ref.~[\onlinecite{weare}], there are several parameter sets which represent exactly identical function $\alpha_{\mbox{\scriptsize HE}} (H)$  of Eq.~({\ref{ha}}). In a general case, it prevents an unambiguous separation of the AHE and ISHE contributions from a fitting of data of Fig. ~\ref{fig:fig2}a,b. However, the case becomes simpler when only AHE, but not ISHE depends on an external parameter. In this case, the current dependency of AHE can be found by the following method. Since the AHE contribution is independent of H, the Hall angle at different current density can be expressed as
\begin{equation}
	\alpha _{HE}^{(I)}\left( H \right) = \alpha _{HE}^{(I =  - 5mA/\mu {m^2})}\left( H \right) + \Delta \alpha _{HE}^{(I)}\; 
	\label{fit}
\end{equation}
where $\Delta \alpha_{\mbox{\scriptsize HE}}^{(I)}$ is independent of H and represents a change of the AHE contribution with current. The value of $\Delta \alpha_{\mbox{\scriptsize HE}}^{(I)}$ is calculated by minimizing a mean square difference between curves of Fig.~\ref{fig:fig2}a. Figure ~\ref{fig:fig2}c shows the result of the minimization. All curves perfectly coincide with each other.

\begin{figure}[b]
\includegraphics[width=8.6cm]{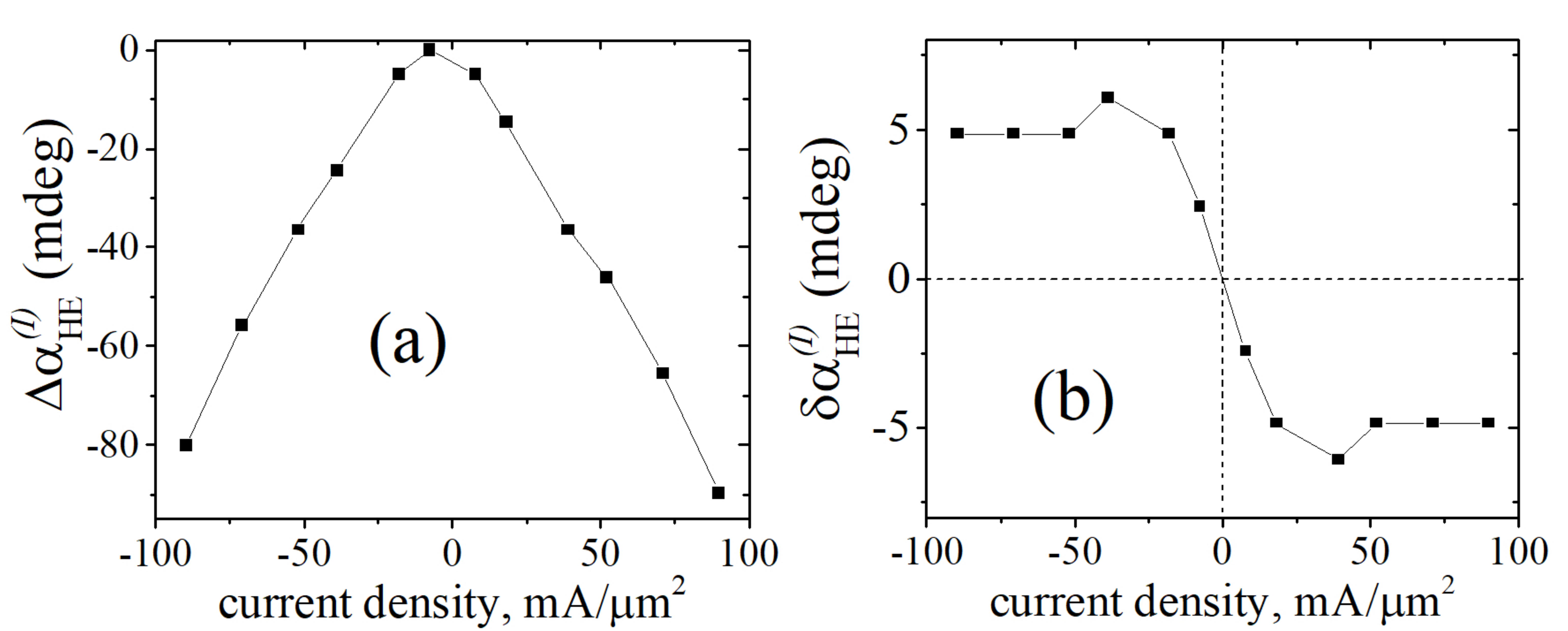}
\caption{\label{fig:fig3}
Dependence of the (a) $\Delta \alpha_{\mbox{\scriptsize HE}}^{(I)}$, which describes the change of the AHE contribution,  and 
(b) polarity-dependent AHE contribution $\delta \alpha_{\mbox{\scriptsize HE}}^{(I)}$ on the current density $I$. 
}
\end{figure}

Figure ~\ref{fig:fig3}a shows current-dependent change of AHE $\alpha_{HE}^{(I)}$. The curve has a parabolic shape, which implies that the AHE change is caused by the Joule-Lenz heating of the nanomagnet proportional to the square of current density $I$. Indeed, the nanomagnet heating up to ${80^o}C$ at $I=90mA/\mu {m^2}$ was confirmed from the measured change of nanomagnet resistance. At this current density, the change of AHE is about 5 percent. The exactly same percentage of change for magnetization is predicted\cite{Gusack} from the Curie-Weiss law for the corresponded temperature change and known FeB Curie temperature $T_c \approx 900 \pm 100$K. Since AHE is linearly proportional to the total spin ${S_d}$ of localized d- electrons, the temperature dependence of ${S_d}$ is perfectly described by the Curie-Weiss law in the studied temperature range.

 Figure ~\ref{fig:fig2}b shows that the ISHE, in contrast to the AHE, is not reduced for the similar 5 percent under the heating and therefore its temperature dependence does not follow the Curie-Weiss law. Indeed, the ISHE is proportional to the total spin ${S_{con}}$ of conduction electrons and therefore to the spin polarization which, in turn, is a function of the spin  pumping and the spin relaxation (See Eqs.~({\ref{p_s}-\ref{d_n}})) and therefore the ISHE temperature dependence is defined by the temperature dependencies of the spin pumping and relaxation rates, but not by the Curie-Weiss law.

Another interesting observed feature is the dependence of the Hall angle on the polarity of current. The parabola of Fig.~\ref{fig:fig3}a is not symmetric with respect to a reversal of  current polarity. The polarity-dependent AHE contribution $\delta \alpha_{\mbox{\scriptsize HE}}^{(I)}$ describes a change of the Hall angle under the same current of different polarity and  is calculated as

\begin{equation}
	\delta \alpha_{\mbox{\scriptsize HE}}^{(I)} = 
	\left( \Delta \alpha_{\mbox{\scriptsize HE}}^{(I)} - \Delta \alpha_{\mbox{\scriptsize HE}}^{(-I)} \right) /2
	\label{asHE}
\end{equation}

Figure ~\ref{fig:fig3}b shows  dependence
$\delta \alpha_{\mbox{\scriptsize HE}}^{(I)}$ on  the current density $I$. Note, the data at a negative and positive current are independent measurements taken at the same parameters. The dependence of  $\delta \alpha_{\mbox{\scriptsize HE}}^{(I)}$ on current is linear with a negative slope at I<40 $mA/\mu {m^2}$ and is saturated at a larger current density.

At first sight, the magnetic and transport properties of the nanomagnet have to be fully symmetric with respect to current reversal. 
The only possible asymmetry can follow from the Spin Hall effect (SHE) because reversal of the current leads to asymmetric spin accumulations on the different sides of wire \cite{Kato2004}.
Then, polarities of the accumulated spins are opposite at the opposite sides of wire and the spin direction is reversed when the current polarity is reversed.  
Strong change of AHE points out to substantial interaction between spins of localized and conduction electrons in the vicinity of interface.
Distribution is different on different interfaces and, hence, this interaction substantially depends on the properties of the interface. 
Since the studied nanomagnet has different materials FeB/Ta and FeB/MgO at opposite interfaces, the spin interaction  should depend on the polarity of spins generated by SHE and therefore the current polarity. The above explains the reason why the total  spins of localized electrons depends on the flow direction of conduction electrons.

\section{Conclusions}

Our study provides strong arguments in favor of the importance of the inverse spin Hall effect
in ferromagnetic metals.  
This conclusion is based on the measurements of the external magnetic field and current dependencies of the 
Hall angle in metallic ferromagnetic samples.  
Adopted mechanisms of ordinary and anomalous Hall effects imply a linear plus constant dependence 
on the external magnetic field whereas our measurements show essentially nonlinear behavior.
We exclude a lot of possible contributions which are not relevant in our nano-sized samples with
perpendicular magnetic anisotropy and find that only inverse spin Hall effect   
can explain the experimental dependence of the Hall effect on the external magnetic field.  We have found that near the room temperature the temperature dependence of the total spin of  localized d- electrons follows the Curie- Weiss law. However, temperature dependence of the total spin of conduction electrons is different and does not follow this law. 
We also observe dependence of the Hall angle of the current polarity. The dependence is due to the Spin Hall effect and the spin interaction of  localized and conduction electrons at interfaces.

\begin{acknowledgments}
We acknowledge stimulating discussions with N. Nagaosa.
This work was supported by JST CREST Grant Number JPMJCR1874, Japan.
\end{acknowledgments}

\bibliography{paperbib}{}

\end{document}